# Near threshold expansion of Feynman diagrams

**E. Mendels**
*Schenkkade 221, 2595 AT The Hague, The Netherlands*

## I. INTRODUCTION

In the usual approach of quantum field theory,[1-4] the successive terms in perturbation theory are represented by diagrams that are given by Feynman rules in momentum space. The most simple Feynman diagram is the zero loop diagram of fig.1.

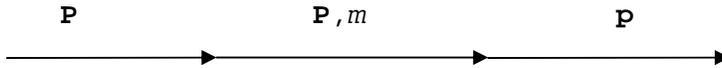

FIG. 1. Zero loop diagram.

Several particles may come in and out, only the sum **p** of their momenta is of interest. The momentum space expression of Fig. 1 is

$$\frac{1}{\mathbf{p}^2 - m^2 + i\varepsilon}. \qquad (1.1)$$

The one loop diagram (fig. 2)

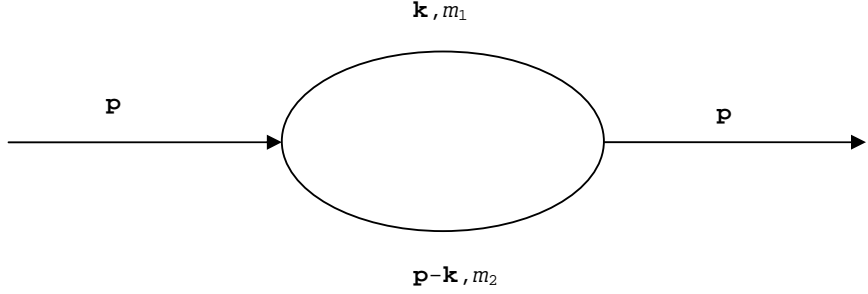

FIG. 2. One loop diagram.

is given by the momentum space expression

$$\int d\mathbf{k} \frac{i}{(\mathbf{k}^2 - m_1^2 + i\varepsilon)\{(\mathbf{p}-\mathbf{k})^2 - m_2^2 + i\varepsilon\}} ,  \qquad (1.2)$$

which is logarithmic divergent in four dimensions. The $i\varepsilon$'s may be transformed away by a Wick rotation, dimensional regularization may be introduced and the internal momentum may be integrated away. Integration over one Feynman parameter is left. In this way, (1.2) becomes[5]

$$(1.2) = \pi^\nu \Gamma(\nu - 2) \int_0^1 dy \{-y(1-y)\mathbf{p}^2 + y m_1^2 + (1-y) m_2^2\}^{\nu-2} . \qquad (1.3)$$

The minus sign in the integrand comes from the inverse Wick rotation at the end of the calculations, which transforms $p$ into $ip$.

The $\nu$ is a continuous variable and (1.3) has to be considered in the limit

$$\varepsilon = \nu - \tfrac{1}{2}n \to 0, \qquad (1.4)$$

where $n$ is the number of dimensions. The logarithmic divergence of (1.2) in four dimensions is visible through the $\Gamma$-function in front,

$$(1.3) = \frac{\pi^2}{\varepsilon} + \pi^2 \int_0^1 dy \ln\{-y(1-y)p^2 + y m_1^2 + (1-y)m_2^2\}. \qquad (1.5)$$

Finally, the integration over $y$ may be done,

$$(1.5) = \frac{\pi^2}{\varepsilon} + \pi^2 \left(\frac{p^2 + m_1^2 - m_2^2}{2p^2}\right) \ln m_1^2 + \pi^2 \left(\frac{p^2 + m_2^2 - m_1^2}{p^2}\right) \ln m_2^2$$

$$- 2\pi^2 + 2\pi^2 \frac{a}{p^2} \operatorname{arctg}\left(\frac{p^2 + m_1^2 - m_2^2}{a}\right) + 2\pi^2 \frac{a}{p^2} \operatorname{arctg}\left(\frac{p^2 + m_2^2 - m_1^2}{a}\right), \qquad (1.6)$$

with

$$a = \sqrt{-(p+m_1+m_2)(p-m_1+m_2)(p+m_1-m_2)(p-m_1-m_2)}. \qquad (1.7)$$

If the absolute value of the argument of the arctangents in (1.6) is smaller than 1, their series expansion is given by

$$\operatorname{arctg}\left(\frac{p^2+m_1^2-m_2^2}{a}\right)=\left(\frac{p^2+m_1^2-m_2^2}{a}\right)-\frac{1}{3}\left(\frac{p^2+m_1^2-m_2^2}{a}\right)^3+\ldots\ldots \quad (1.8a)$$

If it is larger than 1, the series expansion is given by

$$\operatorname{arctg}\left(\frac{p^2+m_1^2-m_2^2}{a}\right)=\frac{\pi}{4}-\left(\frac{a}{p^2+m_1^2-m_2^2}\right)+\frac{1}{3}\left(\frac{a}{p^2+m_1^2-m_2^2}\right)^3-\ldots\ldots. \quad (1.8b)$$

The two-loop diagram (Fig. 3)

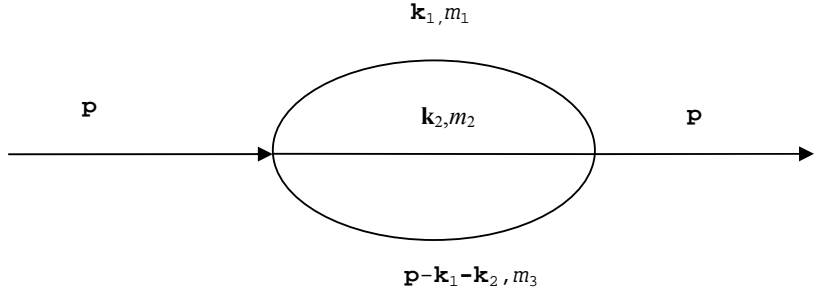

FIG. 3. Two loop diagram.

is represented in momentum space by

$$\int d\mathbf{k}_1 d\mathbf{k}_2 \frac{i^2}{(\mathbf{k}_1^2-m_1^2+i\varepsilon)(\mathbf{k}_2^2-m_2^2+i\varepsilon)\{(\mathbf{p}-\mathbf{k}_1-\mathbf{k}_2)^2-m_3^2+i\varepsilon\}}. \quad (1.9)$$

In four dimensions, it is quadratic divergent and contains subintegrals that are logarithmic divergent. The expression becomes after regularization and integration over the interal momenta,

$$(1.9) = \pi^{2\nu}\Gamma(3-2\nu)\int dy_1 dy_2 \{y_1 y_2 + y_1 y_3 + y_2 y_3\}^{-\nu}$$

$$\times \left(-\frac{y_1 y_2 y_3 p^2}{y_1 y_2 + y_1 y_3 + y_2 y_3} + y_1 m_1^2 + y_2 m_2^2 + y_3 m_3^2\right)^{2\nu-3} \tag{1.10}$$

with

$$y_3 = 1 - y_1 - y_2. \tag{1.11}$$

One of the infinities is visible through the Γ-function in front. Other infinities become visible only after cumbersome partial integrations over the Feynman parameters $y_1$ and $y_2$ and introduction of master integrals to which they are reduced.[6,7]

The cut in the complex $\left(\sum m - p\right)$ plane, causing an imaginary part above threshold, is not clearly visible in (1.2), nor in (1.5). It becomes manifest in (1.6), after having done the $y$ integration. In the two loop case, it is not visible at all, neither in (1.9), nor in (1.10).

Unitarity implies that the imaginary part of diagrams with two vertices and $I$ internal lines should be equal to the phase space integral of $I$ particles. This relation is valid in the zero loop case of Fig. 1,

$$\frac{1}{\mathbf{p}^2 - m^2 + i\varepsilon} = \mathrm{PP}\frac{1}{\mathbf{p}^2 - m^2} - i\pi\delta(\mathbf{p}^2 - m^2). \tag{1.12}$$

In the r.h.s., both a real and an imaginary part is seen. The latter is given by

$$2\Im\left(\frac{1}{\mathbf{p}^2 - m^2 + i\varepsilon}\right) = -2\pi \int d\mathbf{k} \; \delta(\mathbf{p}-\mathbf{k})\theta(k_0)\delta(\mathbf{k}^2 - m^2). \quad (1.13)$$

Indeed, the r.h.s. is the one-particle phase space integral, up to a factor $2\pi$.

In the one loop case of Fig. 2, unitarity is not clear from (1.2) or (1.3). It is seen from (1.8a), which is real if $p < m_1 + m_2$, but if $p > m_1 + m_2$, an imaginary part appears, coming from the $\frac{\pi}{4}$ term in (1.8b):

$$\Im(1.6) = -\frac{\pi^3 \theta(p - m_1 - m_2)\sqrt{(p + m_1 + m_2)(p - m_1 + m_2)(p + m_1 - m_2)(p - m_1 - m_2)}}{p^2}.$$

(1.14)

On the other hand, the two-particle phase space integral in 4 dimensions is given by[8]

$$\int d\mathbf{k}_1 d\mathbf{k}_2 \delta(\mathbf{p} - \mathbf{k}_1 - \mathbf{k}_2)\delta(\mathbf{k}_1^2 - m_1)\delta(\mathbf{k}_2^2 - m_2^2)\theta(k_1^0)\theta(k_2^0)$$

$$= \frac{\pi \theta(p - m_1 - m_2)\sqrt{(p + m_1 + m_2)(p - m_1 + m_2)(p + m_1 - m_2)(p - m_1 - m_2)}}{2p^2}$$

(1.15)

Indeed,

$$2(1.14) = -(2\pi)^2 (1.15) \qquad (1.16)$$

which means unitarity of the one loop diagram in four dimensions.

In the two loop case, the imaginary part of (1.10) is not easily found and if there are near threshold expansions from which it possibly may be derived,[9] it is not clear that it is equal to the three particle phase space integral. On the other hand, if the three particle phase space integral is computed,[10] it is not clear that it equals the imaginary part of (1.10).

To our opinion, it is an unsatisfactory situation that zero loop Feynman rules are given in momentum space, that must be replaced by one loop rules in terms of Feynman parameters that must be replaced once again by new rules that work in the case of more loops. Moreover, it is an unsatisfactory situation that real and imaginary parts are not clearly visible in the usual expressions of Feynman diagrams. Finally, it is an unsatisfactory situation that unitarity is not obviously seen in the usual expressions of Feynman diagrams.

It would be more satisfactory if final rules could be derived, that don't have to be redefined twice. For this reason, it has been proposed[11] to describe Feynman diagrams in terms of configuration space parameters, which are structure conserving and, therefore, do not mix different kinds of infinities.

In the configuration space formalism, internal and external lines of a diagram correspond to Bessel functions with a pure imaginary argument. In the case of sunset diagrams, only one integration variable $x$ is needed. On the other hand, these same Bessel functions appear in the configuration space expression of phase space integrals.[12,13] The *integrands* of sunset diagrams and phase space integrals are equal, only the *ranges* of the $x$ integration are different. By this fact, unitarity is built into the formalism in a natural way.

The **x** space formalism has been extended from sunset diagrams to general Feynman diagrams, which implies integration over several configuration space parameters $\mathbf{x}^j$ and the angles between them.[14]

The purpose of the present paper is to complete the work and to perform all integrations. Rules are derived that immediately give the final result, including cuts, disentangled poles, real and imaginary parts.

The formulas are complicated in the case of a general diagram, summation variables with three and sometimes four indices have to be used. To our opinion, this situation is preferable above a formalism, where infinities, cuts and unitarity are hidden in the integration variables and which too is complicated.

The formalism is general, i.e. it is applicable on diagrams with any number of internal lines, any number of vertices, any number of loops in any dimension and all internal masses may be different. Scalar fields are considered, but introduction of spin does not give complications since spin leads to matrices and derivatives that do not affect the essential structure of the formulas.

## II. SERIES EXPANSION OF SUNSET DIAGRAMS

In order to understand the way of working of the formalism, we consider first dimensionally regularized diagrams with 2 vertices and $I$ internal lines (so called sunset, sunrise or water melon diagrams). They are represented in space by the expression

$$F_2^I \prod \frac{(m^j)^{2b^j}}{\Gamma(\alpha^j)} \int_0^\infty dx\, x^{2\nu-1} i_b(px) k_{b^j}(m^j x) , \qquad (2.1)$$

with notation

$$\prod = \prod_{1 \leq j \leq I} \qquad (2.2)$$

and front factor

$$F_2^I = \frac{(2\pi^{\frac{1}{2}n})^{I+1}}{(2\pi)^n} \ . \tag{2.3}$$

The index *j* distinguishing between the different internal lines is written as superscript. The subscript place has to be reserved for more complicated diagrams.

Formula (2.1) is obtained from the corresponding momentum space expression after a Wick transformation and *n* dimensional Fourier transformations of the propagators according to

$$\int d\mathbf{k}^j \frac{e^{-i\mathbf{k}^j \cdot \mathbf{x}}}{\left(\mathbf{k}^{j^2} + m^{j^2}\right)^{\alpha^j}} = \frac{2\pi^{\frac{1}{2}n} m^{2b^j}}{\Gamma(\alpha^j)} k_{b^j}(m^j x) \ , \tag{2.4}$$

with

$$b^j = \tfrac{1}{2}n - \alpha^j \ . \tag{2.5}$$

Thereupon, one inverse *n* dimensional Fourier transformation has been performed. After integration of an exponential over its angles and an inverse Wick transformation according to

$$\int d\hat{\mathbf{x}} e^{i\mathbf{p}\cdot\mathbf{x}} = 2\pi^{\frac{1}{2}n} j_b(px) \xrightarrow{p \to ip} 2\pi^{\frac{1}{2}n} i_b(px) \ , \tag{2.6}$$

with

$$b = \tfrac{1}{2}n - 1 \ , \tag{2.7}$$

the result (2.1) is obtained.[11] The various factors 2 and $\pi$ in (2.3) are remnants of all these actions.

The functions $i_b(px)$ and $k_{b^j}(m^j x)$ are, up to a factor, modified Bessel functions of the first and third kind,[15] respectively. Details are summarized in appendix A. In the case of integer $b$ and $b^j$, the near zero expansions are given by

$$i_b(z) = j_b(iz) = \sum_{k=0}^{\infty} \frac{1}{k!(b+k)!}\left(\frac{z}{2}\right)^{2k} \qquad (2.8a)$$

$$k_{b^j}(z) = \sum_{0 \leq k^j \leq \infty} K^j \left(\frac{z}{2}\right)^{2k^j} + \sum_{0 \leq k^j \leq \infty} L^j \left(\frac{z}{2}\right)^{2k^j} \ln\left(\frac{z}{2}\right) + \sum_{0 \leq k^j \leq b^j - 1} M^j \left(\frac{z}{2}\right)^{-2b^j + 2k^j} . \qquad (2.8b)$$

$K^j$, $L^j$ and $M^j$ are building stones of the functions $k_{b^j}(z)$; they depend on the summation variable $k^j$ and, according to appendix A, they are given by

$$K^j \equiv K_{b^j}(k^j) = \frac{(-)^{b^j+1}}{k^j!(b^j+k^j)!}\{(\gamma - \tfrac{1}{1} - \tfrac{1}{2} - ... - \tfrac{1}{k^j}) + (\gamma - \tfrac{1}{1} - \tfrac{1}{2} - ... - \tfrac{1}{k^j+b^j})\} \qquad (2.9a)$$

($\gamma = 0{,}5772156649$),

$$L^j \equiv L_{b^j}(k^j) = \frac{(-)^{b^j+1}}{k^j!(b^j+k^j)!} \qquad (2.9b)$$

$$M^j \equiv M_{b^j}(k^j) = \frac{(-)^{k^j}(b^j - k^j - 1)!}{2k^j!} . \qquad (2.9c)$$

As indicated in (2.8b), the summation over $k^j$ is from 0 to $\infty$ if it concerns $K^j$ or $L^j$ and from 0 to $b^j-1$ if it concerns $M^j$.

To compute the integral (2.1), it has to be split according to

$$\int_0^\infty dx = \int_0^X dx + \int_X^\infty dx, \qquad (2.10)$$

where $X$ is arbitrary. In the $\int_X^\infty dx$ integral, the asymptotic expansions (A.6a) for $k_{b^j}(m^j x)$ and (A.6f) for $i_b(px)$ are inserted,

$$\prod \int_X^\infty dx\, x^{2\nu-1} i_b(px) k_{b^j}(m^j x)$$

$$= \sum_{\substack{0 \le k^j < H^j \\ 0 \le k < H}} \prod \frac{\pi^{\frac{I-1}{2}} 2^{2b+2\sum b^c}(b,k)(b^j,k^j) e^{-i\pi(b+\frac{1}{2})}}{(2p)^{b+k+\frac{1}{2}}(2m^j)^{b^j+k^j+\frac{1}{2}}} \int_X^\infty dx\, e^{-\left(\sum m^c + p\right)x} x^{2\nu-(b+k+\frac{1}{2})-\sum(b^c+k^c+\frac{1}{2})-1}$$

$$+ \sum_{\substack{0 \le k^j < H^j \\ 0 \le k < H}} \prod \frac{\pi^{\frac{I-1}{2}} 2^{2b+2\sum b^c}(b,k)(b^j,k^j)}{(2p)^{b+k+\frac{1}{2}}(2m^j)^{b^j+k^j+\frac{1}{2}}} \int_X^\infty dx\, e^{-\left(\sum m^c - p\right)x} x^{2\nu-(b+k+\frac{1}{2})-\sum(b^c+k^c+\frac{1}{2})-1} + \text{rest term},$$

$$(2.11)$$

with notation

$$\sum = \sum_{1 \le c \le I} \qquad (2.12a)$$

$$(b,k) = \frac{\Gamma(b+k+\frac{1}{2})}{k!\,\Gamma(b-k+\frac{1}{2})}. \qquad (2.12b)$$

Though asymptotic expansions of Bessel functions are used,[16-18] it should be remembered that they do not converge in even dimensional spaces[15] (the asymptotic series do converge and even are finite in odd dimensional spaces). Nonetheless, these expansions make sense if their meaning is realized:[19] for a given $H$ and $H^j$ in (2.11), $X$ may be chosen sufficiently large to make the rest term arbitrary small and to make the $X \to \infty$ integrals arbitrary accurate. For this reason, near $\infty$ integrals over the asymptotic series have to be combined with near 0 integrals over expansions that are convergent in the neighbourhood of 0.

The integrals of (2.11) make sense only if the exponentials in the integrand are decreasing, i.e. if

$$\sum m^c - p > 0. \qquad (2.13)$$

By introducing the parameter

$$\Lambda = \left(\sum m^c - p\right) X \qquad (2.14)$$

and choosing $\Lambda$ large with respect to the asymptotic coefficients $(b, H)$ and $(b^j, H^j)$ of (2.12b), both integrals of

(2.11) are small and (2.1) can be computed by doing the integration from 0 to $\Lambda$ only,

$$F_2^I \prod (m^j)^{2b^j} \int_0^\infty dx\, x^{2\nu-1} i_b(px) k_{b^j}(m^j x)$$

$$\cong F_2^I \prod \frac{(\mu^j)^{2b^j}}{\left(\sum m^c - p\right)^{2(\nu-\sum b^c)}} \int_0^\Lambda d\xi\, \xi^{2\nu-1} i_b(\rho\xi) k_{b^j}(\mu^j \xi), \qquad (2.15)$$

with

$$\rho = \frac{p}{\sum m^c - p} \qquad (2.16a)$$

$$\mu^j = \frac{m^j}{\sum m^c - p}. \qquad (2.16b)$$

### III. BELOW THRESHOLD

The Bessel function expansions (2.8) have to be inserted into the r.h.s. of (2.15) and integration is performed with use of appendix D.

In the case of the one loop diagram of Fig. 2, the result is

$$\frac{F_2^2 (\mu^1)^{2b^1} (\mu^2)^{2b^2}}{(m^1 + m^2 - p)^{2(\nu - b^1 - b^2)}} \int_0^\Lambda d\xi \xi^{2\nu - 1} i_b(\rho\xi) k_{b^1}(\mu^1 \xi) k_{b^2}(\mu^2 \xi)$$

$$= \sum_{k,k^1,k^2} \frac{2^{2\nu} F_2^2 \rho^{2k} (\mu^1)^{2(b^1 + k^1)} (\mu^2)^{2(b^2 + k^2)}}{(m^1 + m^2 - p)^{2(\nu - b^1 - b^2)} k!(b+k)!} \tag{3.1}$$

$$\times \left\{ F^{11} \ln\left(\frac{\mu^1 \Lambda}{2}\right) \ln\left(\frac{\mu^2 \Lambda}{2}\right) + F^{10} \ln\left(\frac{\mu^1 \Lambda}{2}\right) + F^{01} \ln\left(\frac{\mu^2 \Lambda}{2}\right) + F^{00} \right\}.$$

The coefficients $F^{11}, F^{10}, F^{01}$ and $F^{00}$ depend on the summation variables $k^j$ and are expressed in terms of the functions $K^j, L^j$ and $M^j$ of (2.9):

$$F^{11} = \frac{L^1 L^2 (\Lambda/2)^N}{N} \tag{3.2a}$$

$$F^{10} = -\frac{L^1 L^2 (\Lambda/2)^N}{(N)^2} + \frac{L^1 K^2 (\Lambda/2)^N}{N} + \frac{L^1 M^2 (\Lambda/2)^{(N - 2b^2)}}{(N - 2b^2)} \tag{3.2b}$$

$$F^{01} = -\frac{L^1 L^2 (\Lambda/2)^N}{(N)^2} + \frac{K^1 L^2 (\Lambda/2)^N}{N} + \frac{M^1 L^2 (\Lambda/2)^{(N - 2b^1)}}{(N - 2b^1)}, \tag{3.2c}$$

$$F^{00} = \frac{L^1 L^2 (\Lambda/2)^N}{(N)^3} - \frac{K^1 L^2 (\Lambda/2)^N}{(N)^2} - \frac{M^1 L^2 (\Lambda/2)^{N - 2b^1}}{(N - 2b^1)^2} - \frac{L^1 K^2 (\Lambda/2)^N}{(N)^2} - \frac{L^1 M^2 (\Lambda/2)^{(N - 2b^2)}}{(N - 2b^2)}$$

$$- \frac{K^1 L^2 (\Lambda/2)^N}{(N)^2} - \frac{M^1 L^2 (\Lambda/2)^{(N - 2b^1)}}{(N - 2b^1)^2} + \frac{K^1 K^2 (\Lambda/2)^N}{N} + \frac{K^1 M^2 (\Lambda/2)^{(N - 2b^2)}}{(N - 2b^2)} + \frac{M^1 K^2 (\Lambda/2)^{N - 2b_1}}{(N - 2b^1)} \tag{3.2d}$$

$$+ \frac{M^1 M^2 (\Lambda/2)^{N - 2b^1 - 2b^2}}{(N - 2b^1 - 2b^2)},$$

with

$$N = 2(\nu + k + k^1 + k^2).\qquad(3.3)$$

In the case of four dimensions ($\nu = 2$) and propagator exponent 1 ($\nu - b^j = 1$), an infinity appears in (3.2d) through the denominator factor $(N - 2b^1 - 2b^2)$. It corresponds to the infinity in (1.5). The formula's (3.1), (3.2) and (3.3) may be generalized to sunset diagrams with $I$ internal lines,

$$\prod \frac{F_2^I(\mu^j)^{2b^j}}{\left(\sum m^c - p\right)^{2(\nu - \sum b^c)}} \int_0^\Lambda d\xi \xi^{2\nu-1} i_b(\rho\xi) k_{b^j}(\mu^j\xi)$$

$$= \sum_{k,k^j} \prod \frac{2^{2\nu} F_2^I \rho^{2k}(\mu^j)^{2(b^j+k^j)}}{\left(\sum m^c - p\right)^{2(\nu-\sum b^c)} k!(b+k)!} \sum_{0 \le f^j \le 1} F^{f^1...f^j}\left\{\ln\left(\frac{\mu^j\Lambda}{2}\right)\right\}^{f^j},$$

(3.4)

with

$$F^{f^1...f^j} = \sum_{f^j \le g^j \le h^j \le 1} \prod \frac{(-)^G G! (K^j)^{(1-h^j)(1-g^j)} (L^j)^{g^j} (M^j)^{(1-g^j)h^j}}{\left\{N - 2\sum(1-g^c)h^c b^c\right\}^{1+G}} \left(\frac{\Lambda}{2}\right)^{N-2\sum(1-g^c)h^c b^c},$$

(3.5)

where

$$G = \sum(g^c - f^c)\qquad(3.6a)$$

$$N = 2(\nu + k + \sum_{1 \le c \le I} k^c).\qquad(3.6b)$$

To (3.4) has to be added the $\Lambda \to \infty$ integrals (2.11) which may help convergence and may be computed by means of the formulas of appendix C. The latter integrals are arbitrary small for sufficiently large $\Lambda$.

## IV. ABOVE THRESHOLD

The threshold at $p = \sum m^c$ appears through the logarithmic factors in (3.1) and (3.4). It is a direct consequence of the requirement (2.13) that the asymptotic integrands (2.11) should decrease exponentially.

Analytical continuation to above threshold, i.e. to

$$p > \sum m , \qquad (4.1)$$

is obtained AFTER replacement of $\left(\sum m - p\right)$ by $\left(p - \sum m\right)$ in (3.1) and (3.4). Terms that are even in $\left(\sum m - p\right)$ remain unchanged by this replacement, odd terms transform into their opposite and a term $i\pi$ is added to a logarithmic factor,

$$\left(\sum m - p\right)^{2k} \to \left(p - \sum m\right)^{2k} \qquad (4.2a)$$

$$\left(\sum m - p\right)^{2k+1} \to -\left(p - \sum m\right)^{2k+1} \qquad (4.2b)$$

$$\ln\left(\sum m - p\right) \to \ln\left(p - \sum m\right) + i\pi . \qquad (4.2c)$$

Imaginary terms appear by these transformations through the factor

$$\sum_{0 \leq f^i \leq 1} F^{f^1, f^2} \left\{ \ln\left(\frac{\mu^1 \Lambda}{2}\right) \right\}^{f^1} \left\{ \ln\left(\frac{\mu^2 \Lambda}{2}\right) \right\}^{f^2}.$$

In the one loop case,

$$F^{11} \ln\left(\frac{\mu^1 \Lambda}{2}\right) \ln\left(\frac{\mu^2 \Lambda}{2}\right) + F^{10} \ln\left(\frac{\mu^1 \Lambda}{2}\right) + F^{01} \ln\left(\frac{\mu^2 \Lambda}{2}\right) + F^{00}$$

$$\rightarrow F^{1,1} \ln\left(\frac{\mu^1 \Lambda}{2}\right) \ln\left(\frac{\mu^2 \Lambda}{2}\right) - \pi^2 F^{1,1} + F^{1,0} \ln\left(\frac{\mu^1 \Lambda}{2}\right) + F^{0,1} \ln\left(\frac{\mu^2 \Lambda}{2}\right) + F^{0,0} \qquad (4.3)$$

$$+ i\pi \left\{ F^{1,1} \ln\left(\frac{\mu^1 \Lambda}{2}\right) + F^{1,1} \ln\left(\frac{\mu^2 \Lambda}{2}\right) + F^{1,0} + F^{0,1} \right\}.$$

## V. THE GENERAL FEYNMAN DIAGRAM

A general Feynman diagram, with $V$ vertices and $I$ internal lines, is written in terms of configuration space parameters as[14]

$$\frac{(2\pi^{\frac{1}{2}n})^I}{(2\pi)^{n(V-1)}} \prod \frac{(m_i^j)^{2b_i^j} (m_{i',i}^j)^{2b_{i',i}^j}}{\Gamma(\alpha_i^j) \Gamma(\alpha_{i',i}^j)} \int d\mathbf{x}_i e^{i\mathbf{p}_i \cdot \mathbf{x}_i} k_{b_i^j}(m_i^j x_i) k_{b_{i',i}^j}\left(m_{i',i}^j \,|\, \mathbf{x}_i - \mathbf{x}_{i'}|\right) \qquad (5.1)$$

with notation

$$\prod = \prod_{\substack{1 \leq i \leq V-1 \\ 1 \leq i' < i \\ j}} . \qquad (5.2)$$

Internal lines and their corresponding masses wear three indices: the two subscripts $i',i$ indicate the vertices that they are connecting, the superscript $j$ counts the lines between these vertices.

The number of integration vectors $\mathbf{x}_i$ is $V-1$, one less than the number of vertices. The external momenta are given by $V-1$ vectors $\mathbf{p}_i$, the momentum $\mathbf{p}_V$ is given by momentum conservation:

$$\sum_{1 \leq i \leq V} \mathbf{p}_i = 0 . \qquad (5.3)$$

The elaboration of (5.1) goes along the same lines as in the case of section 2. It will turn out that the final result is a generalization of the result (3.4) for sunset diagrams.

The integrand of (5.1) contains both functions like $k_{b_i}(m_i x_i)$, with a simple argument, and functions like $k_{b_{i',i}}(m_{i',i} |\mathbf{x}_i - \mathbf{x}_{i'}|)$, with a composed argument. Because of the latter functions, the integration range of (5.1) has to be divided into $V!$ subregions. $(V-1)!$ of them are of the kind

$$0 < x_1 < x_2 < ... < x_{V-1} < X \quad + \text{ permutations}, \tag{5.4a}$$

with arbitrary $X$,

and $V!-(V-1)!$ subregions are of the kind

$$0 < x_1 < x_2 < .. < X < .. < x_{V-1} < \infty . \tag{5.4b}$$

In the subregions of (5.4b), the functions $k_{b_{i',i}}$ with a composed argument are split according to (A.8) and the integral (5.1) contains a sum of products of three factors:

$$(5.1) = F_V^I \sum \prod \frac{(m_i^j)^{2b_i^j} (m_{i',i}^j)^{2b_{i',i}^j}}{\Gamma(\alpha_i^j)\Gamma(\alpha_{i',i}^j)} A(r_{i',i}^j) R_1(r_{i',i}^j) R_2(r_{i',i}^j) , \tag{5.5}$$

with front factor

$$F_V^I = \frac{(2\pi^{\frac{1}{2}n})^{I+V-1}}{(2\pi)^{n(V-1)}} . \tag{5.6}$$

$A(r_{i',i}^j)$ is an angular factor given in (B.8b), $R_1(r_{i',i}^j)$ and $R_2(r_{i',i}^j)$ are radial factors given by (C.1a) and (C.1b).

The radial integrals are meaningful only if the integrands are decreasing exponentials. From the formula's of appendix C, it is seen that the latter condition is satisfied below all thresholds, i.e. if

$$\sum m - \sum p \equiv \min_V \left\{ \sum_{a \in V, b \notin V} \left( \sum_c m_{a,b}^c - p_b \right) \right\} > 0 , \quad (5.7)$$

where $V$ is a set of vertices and $\min_V$ is the minimum over all possible sets. Under the condition (5.7), integrals in the subregions (5.4b) can be made arbitrary small if

$$\Lambda \equiv \left( \sum m - \sum p \right) X \quad (5.8)$$

is chosen sufficiently large.

The non disappearing integrals in the subregions (5.4a) are computed by splitting the composed functions $k_{b_{i',i}}(m_{i',i} | \mathbf{x}_i - \mathbf{x}_{i'} |)$ according to (A.11), which leads to a simpler formula than (A.8). (5.1) is written in these subregions as the sum over products of an angular and a radial factor according to

$$(5.1) \cong F_V^I \sum \prod \frac{r_{i',i}!}{\Gamma(\alpha_i^j)\Gamma(\alpha_{i',i}^j) r_{i',i}^j} A(q_i; r_{i',i}) R(q^i; r_{i',i}), \quad (5.9)$$

with

$$r_{i',i} = \sum_c r_{i',i}^c , \quad (5.10a)$$

$$(2\pi^{\frac{1}{2}n})^{V-1} A(q_i, r_{i',i}) = \prod \int d\hat{\mathbf{x}}_i \frac{(2\hat{\mathbf{p}}_i \cdot \hat{\mathbf{x}}_i)^{q_i}}{q_i!} \frac{(2\hat{\mathbf{x}}_{i'} \cdot \hat{\mathbf{x}}_i)^{r_{i',i}}}{r_{i',i}!} \qquad (5.10b)$$

and

$$R(q_i; r_{i',i}^j) \equiv \sum_{l_{i',i}^j=0}^{\infty} \prod \frac{2^{2\nu(V-1)}(-)^{l_{i',i}^j}}{l_{i',i}^j!} p_i^{q_i} (m_i^j)^{2b_i^j} (m_{i',i}^j)^{2(b_{i',i}^j + l_{i',i}^j + r_{i',i}^j)}$$

$$\times \int_{0<x_1<\ldots<x_{V-1}<X} d\frac{x_i}{2} \left(\frac{x_i}{2}\right)^{2\nu-1+q_i+2\sum l_{i,a}^c + \sum r_{a,i} + \sum r_{i,a}} k_{b_i^j}(m_i^j x_i) k_{b_{i',i}^j + l_{i',i}^j + r_{i',i}^j}(m_{i',i}^j x_i) \qquad (5.10c)$$

+permutations.

Summation $\sum$ in the exponent is defined in (B.5).

Integral (5.10b) is computed in (B.4b) and (5.10c) is computed by insertion of (A.3c). The final result is a generalization of the result (3.4) for sunset diagrams:

$$(5.1) \cong \frac{2^{2\nu(V-1)} F_V^I}{\left(\sum m - \sum p\right)^{2\{\nu(V-1) - \sum b_a^c - \sum b_{a,b}^c\}}} \sum_{k,l,q,r} \prod \frac{r_{i',i}!}{\Gamma(\alpha_i^j)\Gamma(\alpha_{i',i}^j) r_{i',i}^j} A(q_i; r_{i',i}) \rho_i^{q_i} \qquad (5.11)$$

$$\times \frac{(-)^{l_{i',i}^j}}{l_{i',i}^j} (\mu_i^j)^{2(b_i^j + k_i^j)} (\mu_{i',i}^j)^{2(b_{i',i}^j + l_{i',i}^j + r_{i',i}^j + k_{i',i}^j)} \sum_{f=0}^{1} F^{f_i^j, f_{i',i}^j} \left\{\ln\left(\frac{\mu_i^j \Lambda}{2}\right)\right\}^{f_i^j} \left\{\ln\left(\frac{\mu_{i',i}^j \Lambda}{2}\right)\right\}^{f_{i',i}^j},$$

with

$$\rho_i = \frac{p_i}{\left(\sum m - \sum p\right)}, \quad \mu_i^j = \frac{m_i^j}{\left(\sum m - \sum p\right)}, \quad \mu_{i',i}^j = \frac{m_{i',i}^j}{\left(\sum m - \sum p\right)} \quad (5.12a)$$

and

$$F^{f_i^j, f_{i',i}^j} = \sum_{f \leq g_{;V-2} \ldots \leq g_{;1} \leq g \leq h \leq 1} \prod \frac{(-)^{G_i} G_i!}{\left(\sum_{1 \leq a \leq i} N_a\right)^{1+G_i}} (K_i^j)^{(1-g_i^j)(1-h_i^j)} (L_i^j)^{g_i^j} (M_i^j)^{(1-g_i^j)h_i^j} (K_{i',i}^j)^{(1-g_{i',i}^j)(1-h_{i',i}^j)} (L_{i',i}^j)^{g_{i',i}^j} (M_{i',i}^j)^{(1-g_{i',i}^j)h_{i',i}^j}$$

$$\times \left(\frac{\Lambda}{2}\right)^{\sum_{1 \leq a \leq V-1} N_a} + \text{permutations}.$$

(5.12b)

$F^{f_i^j, f_{i',i}^j}$ depends on the summation variables $k, l, q, r$ through the factors $K_i^j, L_i^j, M_i^j$, through the factors

$$K_{i',i}^j \equiv K_{b_{i',i}^j + l_{i',i}^j + r_{i',i}^j}(k_{i',i}^j), \quad L_{i',i}^j \equiv L_{b_{i',i}^j + l_{i',i}^j + r_{i',i}^j}(k_{i',i}^j), \quad M_{i',i}^j \equiv M_{b_{i',i}^j + l_{i',i}^j + r_{i',i}^j}(k_{i',i}^j), \quad (5.13)$$

all defined according to (2.9), and through the factors $\sum_{1 \leq a \leq i} N_a$ with

$$N_i = 2v + q_i + 2\sum k_i^c + 2\sum k_{a,i}^c + 2\sum l_{i,a}^c + \sum r_{a,i}^c + \sum r_{i,a}^c$$
$$- 2\sum h_i^c(1-g_i^c)b_i^c - \sum h_{a,i}^c(1-g_{a,i}^c)(b_{a,i}^c + l_{a,i}^c + r_{a,i}^c),$$

(5.14)

where summation $\sum$ is according to the convention of (B.5).

$G_i$ is computed in appendix D:

$$G_i = \sum_{\substack{1 \leq a \leq i \\ c}} (g^c_{a;i-1} - g^c_{a;i}) + \sum_{\substack{1 \leq a' < a \leq i \\ c}} (g^c_{a',a;i-1} - g^c_{a',a;i}), \quad (5.15)$$

where $g^j_i$, $g^j_{i',i}$, $g^j_{i;i''}$, $g^j_{i',i;i''}$, $f^j_i$, and $f^j_{i',i}$ ($i' < i \leq i'' < V-1$) are summation variables running from 0 to 1 and

$$g^j_{i;i-1} \equiv g^j_i \quad (5.16a)$$

$$g^j_{i',i;i-1} \equiv g^j_{i',i} \quad (5.16b)$$

$$g^j_{i;V-1} \equiv f^j_i \quad (5.16c)$$

$$g^j_{i',i;V-1} \equiv f^j_{i',i}. \quad (5.16d)$$

$\sum_{f \leq g_{;V-2} \ldots \leq g_{;1} \leq g \leq h \leq 1}$ in (5.12b) means summation under the conditions

$$f^j_i \leq g^j_{i;V-2} \leq \ldots \leq g^j_{i;i} \leq g^j_i \leq h^j_i \leq 1 \quad (5.17a)$$

$$f^j_{i',i} \leq g^j_{i',i;V-2} \leq \ldots \leq g^j_{i',i;i} \leq g^j_{i',i} \leq h^j_{i',i} \leq 1 \quad (5.17b)$$

$\sum_k$ means summation over all $k_i^j$ and $k_{i',i}^j$ from 0 to $\infty$, if they are appearing through factors $K$ and $L$, the summation is from 0 to $b_i^j - 1$ and $b_i^j + l_i^j + r_i^j - 1$, respectively, if they are appearing through a factor $M$.

(5.11) is a generalization of formula (3.4) for sunset diagrams. It is a sum of terms that have factors $K$, $L$ and $M$, defined in (2.8) and (2.9), in the numerator. In stead of a power of one denominator factor $N$ in (3.4), (5.10c) leads to $V-1$ denominator factors $\sum_{a=1}^{i} N_a$ $(i = 1, 2, ..., V-1)$, which all, in turn, may have higher powers, caused by integration over logarithms. The various terms of (5.12b) may be described by the following features:

- The first term consists of $I$ factors $L$ in the numerator. The exponent of $x_i$ in (5.10c) becomes, after integration, a factor $\sum_{a=1}^{i} N_a$, in the denominator. The factor in front is $0!$.
- Integration of terms with factors $L$ gives, because of the logarithm of (2.8b), both a term with a logarithm and a term without logarithm.

- In the latter case, the power of the corresponding factors $\sum_{a=1}^{i} N_a$ increases by 1, as well as the argument of the faculty in front, and the sign of the term changes.
- Further terms are obtained after replacement of factors $L$ by $K$, $M$ or combinations in terms with an increased power of $\sum_{a=1}^{i} N_a$. Each replacement lowers the power of $\sum_{a=1}^{i} N_a$ by 1 as well as the argument of the faculty in front, and changes the sign of the term.
- A factor $M$ in the numerator implies addition of a term $-2b$ or $-2(b+l+r)$ to the corresponding $N_a$ in the denominator. If two or more of these terms are added, poles in the complex $v$ plane may appear.

The integrals in the regions (5.4b), which may be computed by means of the formulas of appendix C, have to be added to (5.11) to help convergence. The latter integrals are arbitrary small for sufficiently large $\Lambda$.

Analytical continuation from the region (5.7) to other regions of the complex $\left(\sum m - \sum p\right)$ plane is obtained according to the rules of (4.2).

As an example, we demonstrate a diagram with three vertices and five internal lines according to Fig. 4.

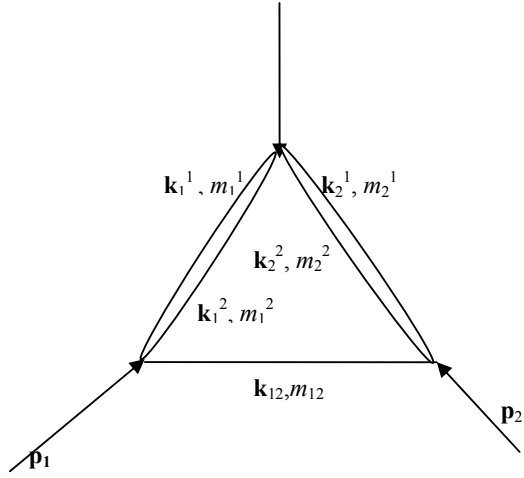

FIG. 4. A diagram with three vertices and five internal lines.

The expression is

$$\frac{(2\pi^{1/2n})^5}{(2\pi)^{2n}} \frac{(m_1^1)^{2b_1^1}(m_1^2)^{2b_1^2}(m_2^1)^{2b_2^1}(m_2^2)^{2b_2^2}(m_{1,2})^{2b_{1,2}}}{\Gamma(\alpha_1^1)\Gamma(\alpha_1^2)\Gamma(\alpha_2^1)\Gamma(\alpha_2^2)\Gamma(\alpha_{1,2})}$$

$$\times \int d\mathbf{x}_1 d\mathbf{x}_2 e^{i\mathbf{p}_1\cdot\mathbf{x}_1} e^{i\mathbf{p}_2\cdot\mathbf{x}_2} k_{b_1^1}(m_1^1 x_1) k_{b_1^2}(m_1^2 x_1) k_{b_2^1}(m_2^1 x_2) k_{b_2^2}(m_2^2 x_2) k_{b_{1,2}}\left(m_{1,2}|\mathbf{x}_1 - \mathbf{x}_2|\right)$$

(5.18)

and the region of the radial integrations is divided into subregions $0 < x_1 < x_2 < X$, $0 < x_1 < X < x_2$, $0 < X < x_1 < x_2$ and permutations of $x_1$ and $x_2$. $\left(\sum m - \sum p\right)$ is given by

$$\left(\sum m - \sum p\right) \equiv \min \left\{ \begin{array}{l} (m_1^1 + m_1^2 + m_{1,2} - p_1), \\ \\ (m_2^1 + m_2^2 + m_{12} - p_2), (m_1^1 + m_1^2 + m_2^1 + m_2^2 - p_1 - p_2) \end{array} \right\}. \quad (5.19)$$

The contribution of the subregions $0 < x_1 < X < x_2$, $0 < X < x_1 < x_2$ and permutations can be made arbitrary small by choosing

$$\Lambda = \left(\sum m - \sum p\right) X \quad (5.20)$$

sufficiently large. In this case, only the two regions $0 < x_1 < x_2 < X$ and $0 < x_2 < x_1 < X$ contribute. In the latter subregions, (5.18) is factorized into an angular and a radial part:

$$(5.18) \cong \frac{(2\pi^{\frac{1}{2}n})^7}{(2\pi)^{2n} \Gamma(\alpha_1^1)\Gamma(\alpha_1^2)\Gamma(\alpha_2^1)\Gamma(\alpha_2^2)\Gamma(\alpha_{12})} A(q_1, q_2, r_{12}) R(q_1, q_2, r_{12}) \quad (5.21)$$

with

$$A(q_1, q_2, r_{12}) = \frac{1}{(2\pi^{\frac{1}{2}n})^2} \int d\hat{\mathbf{x}}_1 d\hat{\mathbf{x}}_2 \frac{(2\hat{\mathbf{p}}_1 \cdot \hat{\mathbf{x}}_1)^{q_1}}{q_1!} \frac{(2\hat{\mathbf{p}}_2 \cdot \hat{\mathbf{x}}_2)^{q_2}}{q_2!} \frac{(2\hat{\mathbf{x}}_1 \cdot \hat{\mathbf{x}}_2)^{r_{1,2}}}{r_{1,2}!}$$

$$= \sum_{q_{1,2}, p_{1,2}}{}' \frac{(2\hat{\mathbf{p}}_1 \cdot \hat{\mathbf{p}}_2)^{p_{1,2}}}{\frac{1}{2}(n+q_1+r_{1,2})! \frac{1}{2}(n+q_2+q_{1,2})! \frac{1}{2}(q_1-q_{1,2})! \frac{1}{2}(q_2-p_{1,2})! \frac{1}{2}(r_{1,2}-q_{1,2})! \frac{1}{2}(q_{1,2}-p_{1,2})! p_{1,2}!}$$

(5.22a)

($\sum'$ means summation over $q_{1,2}$ and $p_{1,2}$ under the condition that the arguments of the faculties in the denominator are integers) and

$$R(q_1,q_2,r_{12}) = \sum_{l_{1,2}} \frac{2^{4v} p_1^{q_1} p_2^{q_2} (m_1^1)^{2b_1^1} (m_1^2)^{2b_1^2} (m_2^1)^{2b_2^1} (m_2^2)^{2b_2^2} (-)^{l_{1,2}} (m_{1,2})^{2(b_{1,2}+l_{1,2}+r_{1,2})}}{l_{1,2}!}$$

$$\int_{0<x_1<x_2<X} d\frac{x_1}{2} d\frac{x_2}{2} \left(\frac{x_1}{2}\right)^{2v+q_1+2l_{1,2}+r_{1,2}-1} \left(\frac{x_2}{2}\right)^{2v+q_2+r_{1,2}-1}$$

$$\times k_{b_1^1}(m_1^1 x_1) k_{b_1^2}(m_1^2 x_1) k_{b_2^1}(m_2^1 x_2) k_{b_1^2}(m_2^2 x_2) k_{b_{1,2}+l_{1,2}+r_{1,2}}(m_{1,2} x_2) + \text{permutation}.$$

(5.22b)

The r.h.s. of (5.22b) is computed with use of appendix D. The final result is

$$(5.18) \cong \sum \frac{2^{4\nu} F_3^5 A(q_1, q_2, r_{1,2}) \rho_1^{q_1} \rho_2^{q_2} (\mu_1^1)^{2(b_1^1+k_1^1)} (\mu_1^2)^{2(b_1^2+k_1^2)} (\mu_2^1)^{2(b_2^1+k_2^1)} (\mu_2^2)^{2(b_2^2+k_2^2)} (-)^{l_{1,2}} (\mu_{1,2})^{2(b_{1,2}+l_{1,2}+r_{1,2}+k_{1,2})}}{\left(\sum m - \sum p\right)^{4\nu - 2(b_1^1+b_1^2+b_2^1+b_2^2+b_{1,2})} l_{1,2}!}$$

$$\times \left\{ \begin{aligned}
&\left\{ \frac{L_1^1 L_1^2 L_2^1 L_2^2 L_{1,2}}{N_1 (N_1 + N_2)} \right\} \ln\left(\frac{\mu_1^1 \Lambda}{2}\right) \ln\left(\frac{\mu_1^2 \Lambda}{2}\right) \ln\left(\frac{\mu_2^1 \Lambda}{2}\right) \ln\left(\frac{\mu_2^2 \Lambda}{2}\right) \ln\left(\frac{\mu_{1,2} \Lambda}{2}\right) \left(\frac{\Lambda}{2}\right)^{N_1+N_2} \\
&+ \left\{ \begin{aligned} &-\frac{L_1^1 L_1^2 L_2^1 L_2^2 L_{1,2}}{N_1^2 (N_1+N_2)} - \frac{L_1^1 L_1^2 L_2^1 L_2^2 L_{1,2}}{N_1 (N_1+N_2)^2} \\ &+\frac{K_1^1 L_1^2 L_2^1 L_2^2 L_{1,2}}{N_1 (N_1+N_2)} + \frac{M_1^1 L_1^2 L_2^1 L_2^2 L_{1,2}}{(N_1 - 2b_1^1)(N_1+N_2)} + \ldots \end{aligned} \right\} \ln\left(\frac{\mu_1^2 \Lambda}{2}\right) \ln\left(\frac{\mu_2^1 \Lambda}{2}\right) \ln\left(\frac{\mu_2^2 \Lambda}{2}\right) \ln\left(\frac{\mu_{1,2} \Lambda}{2}\right) \left(\frac{\Lambda}{2}\right)^{N_1+N_2} \\
&+ \left\{ -\frac{L_1^1 L_1^2 L_2^1 L_2^2 L_{1,2}}{N_1 (N_1+N_2)^2} + \frac{L_1^1 L_1^2 K_2^1 L_2^2 L_{1,2}}{N_1 (N_1+N_2)} + \ldots \right\} \ln\left(\frac{\mu_1^1 \Lambda}{2}\right) \ln\left(\frac{\mu_1^2 \Lambda}{2}\right) \ln\left(\frac{\mu_2^2 \Lambda}{2}\right) \ln\left(\frac{\mu_{1,2} \Lambda}{2}\right) \left(\frac{\Lambda}{2}\right)^{N_1+N_2} \\
&+ \left\{ \frac{L_1^1 L_1^2 M_2^1 L_2^2 L_{1,2}}{N_1 (N_1+N_2 - 2b_2^1)} + \ldots \right\} \ln\left(\frac{\mu_1^1 \Lambda}{2}\right) \ln\left(\frac{\mu_1^2 \Lambda}{2}\right) \ln\left(\frac{\mu_2^2 \Lambda}{2}\right) \ln\left(\frac{\mu_{1,2} \Lambda}{2}\right) \left(\frac{\Lambda}{2}\right)^{N_1+N_2-2b_2^1} \\
&+ \left\{ \frac{2! L_1^1 L_1^2 L_2^1 L_2^2 L_{1,2}}{N_1 (N_1+N_2)^3} - \frac{L_1^1 L_1^2 L_2^1 K_2^2 L_{1,2}}{N_1 (N_1+N_2)^2} + \ldots \right\} \ln\left(\frac{\mu_1^1 \Lambda}{2}\right) \ln\left(\frac{\mu_1^2 \Lambda}{2}\right) \ln\left(\frac{\mu_2^1 \Lambda}{2}\right) \left(\frac{\Lambda}{2}\right)^{N_1+N_2} \\
&+ \left\{ -\frac{L_1^1 L_1^2 L_2^1 K_2^2 M_{1,2}}{N_1 \{N_1 + N_2 - 2(b_{1,2}+l_{1,2}+r_{1,2})\}^2} + \ldots \right\} \ln\left(\frac{\mu_1^1 \Lambda}{2}\right) \ln\left(\frac{\mu_1^2 \Lambda}{2}\right) \ln\left(\frac{\mu_2^1 \Lambda}{2}\right) \left(\frac{\Lambda}{2}\right)^{N_1+N_2-2(b_{1,2}+l_{1,2}+r_{1,2})} \\
&+ \left\{ -\frac{L_1^1 L_1^2 L_2^1 M_2^2 M_{1,2}}{N_1 \{N_1 + N_2 - 2(b_2^2+b_{1,2}+l_{1,2}+r_{1,2})\}^2} + \ldots \right\} \ln\left(\frac{\mu_1^1 \Lambda}{2}\right) \ln\left(\frac{\mu_1^2 \Lambda}{2}\right) \ln\left(\frac{\mu_2^1 \Lambda}{2}\right) \left(\frac{\Lambda}{2}\right)^{N_1+N_2-2(b_2^2+b_{1,2}+l_{1,2}+r_{1,2})}
\end{aligned} \right\}$$

+ many similar terms + permutations,

(5.23)

with

$$N_1 = 2\nu + q_1 + 2k_1^1 + 2k_1^2 + 2l_{1,2} + r_{1,2} \qquad (5.24a)$$

$$N_2 = 2\nu + q_2 + 2k_2^1 + 2k_2^2 + 2k_{1,2} + r_{1,2} . \qquad (5.24b)$$

## VI. UNITARITY

The S matrix has to satisfy the unitarity relation

$$S^\dagger S = I. \qquad (6.1)$$

In terms of the T-matrix, which is defined by

$$S = I - i\delta(\Sigma \mathbf{p})T, \qquad (6.2)$$

(6.1) is written as

$$2\Im(T) \equiv i(T^\dagger - T) = -\delta(\Sigma \mathbf{p})T^\dagger T. \qquad (6.3)$$

In the case of sunset diagrams, their imaginary part has to be equal to the integral over positive energy states with momenta that are on the mass shell, which is the phase space integral.

The $n$ dimensional Laplace transform of $\theta(k^0)\delta(\mathbf{k}^2 - m^2)$ is found by using the integral representation (A.14):[13]

$$\int_0^\infty d\mathbf{k} e^{-\mathbf{x}\cdot\mathbf{k}} \theta(k^0)\delta(\mathbf{k}^2 - m^2) = \frac{\pi^{\frac{n-1}{2}}}{\Gamma\left(\frac{n-1}{2}\right)} \int_m^\infty dk^0 e^{-xk^0} (k^{0^2} - m^2)^{\frac{n-3}{2}}$$

$$(6.4)$$

$$= \frac{\pi^{\frac{n-1}{2}} (m^2)^{\frac{n}{2}-1}}{\Gamma\left(\frac{n-1}{2}\right)} \int_1^\infty dt e^{-mxt} (t^2 - 1)^{\frac{n-3}{2}} = \pi^{\frac{n-1}{2}}(m^2)^{\frac{n}{2}-1} k_{\frac{n}{2}-1}(mx).$$

Applying (6.4) on a product of factors $\theta(k^0)\delta(\mathbf{k}^2 - m^2)$ and taking afterwards the inverse $n$ dimensional Laplace transformation, which is an inverse K transform according to (A.15), it follows that[12,13]

$$(2\pi)^I \prod \int d\mathbf{k}^j \delta(\mathbf{p} - \sum \mathbf{k}^c)\theta(k^j{}_0)\delta(\mathbf{k}^{j2} - m^{j2}) = \frac{F_2^I}{i}\prod (m^j)^{2b^j} \int_{c-i\infty}^{c+i\infty} dx\, x^{n-1} i_b(px) k_{b^j}(m^j x) ,$$

(6.5)

with

$$b = b^j = \tfrac{1}{2}n - 1 \qquad (6.6a)$$

$$c > 0 , \qquad (6.6b)$$

and $F_2^I$ is given in (2.3). The l.h.s. of (6.5) is, up to a factor $(2\pi)^I$, the $I$ particle phase space integral.

We remark that (6.5) is obtained by Laplace transformations followed by an $n$ dimensional inverse Laplace transformation in same way as (2.1) is obtained from the momentum space representation by Fourier transformations followed by an $n$ dimensional inverse Fourier transformation. The integrands of (6.5) and (2.1) are equal, the integrals differ by the path of integration.

In the case $p < \sum m$, it is seen that (6.5) is 0, by closing the contour to the right of the complex $x$ plane and by remarking that the integrand vanishes towards ∞ (fig. 5).

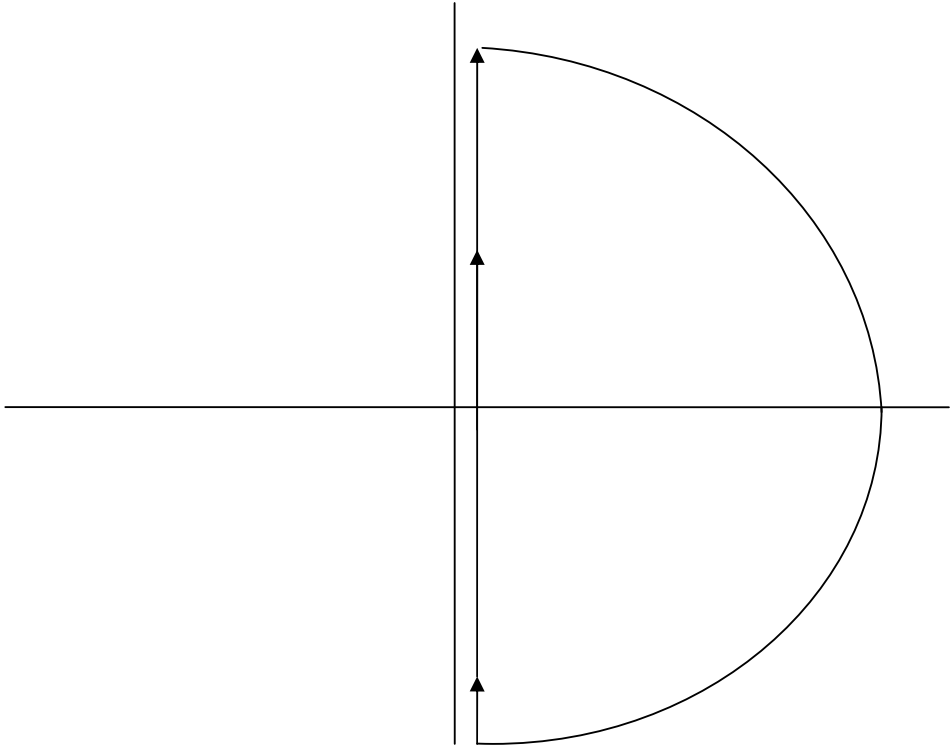

Fig. 5. Integration path to the right.

In the case $p > \sum m$, the integral does not vanish near infinity towards the left hand side of the complex plane, since the decreasing asymptotic behaviour of the integrand is restricted by the values of the argument of the integration variable, as indicated in (A.6). The integrand vanishes near ∞ only if crossing the negative axis is avoided. Hence, the path of integration has to be closed according to fig. 6.

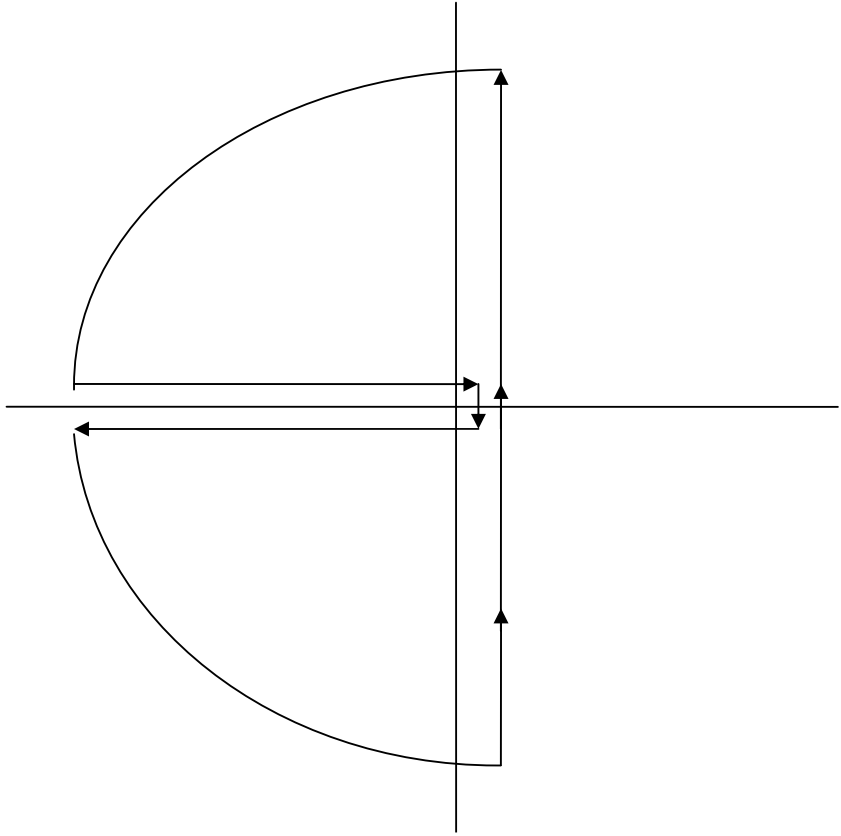

Fig. 6. Integration path to the left.

The contribution from the near zero part and that of the near infinity circle are both 0. The contribution along the negative real axis from -∞→-0 is equal to the complex conjugate of the -0→-∞ integration.

Hence, we conclude that

$$(6.5) = -F_2^I \prod (m^j)^{2b^j} \int_0^\infty dx\, x^{n-1} 2\Im\left\{ \prod i_{\frac{n}{2}-1}(px) k_{\frac{n}{2}-1}(m^j x) \right\}. \qquad (6.7)$$

After interchanging integral and $\Im$ operation, the unitarity relation

$$2\Im(2.1) = -(2\pi)^I \prod \int d\mathbf{k}^j \delta(\mathbf{p} - \sum \mathbf{k}^j) \theta(k^j{}_0) \delta(\mathbf{k}^{j2} - m^{j2}) \qquad (6.8)$$

of a sunset diagram is obtained. It is in agreement with (1.13) and (1.16).

## VII. CONVERGENCE

Expansions with many summation variables occurred in the previous sections and convergence of these series has to be discussed.

In the used formulas, faculties of summation variables appear in the denominator, which helps convergence. Also factors like $(-)^k$ and $(-)^l$ may promote convergence.

A point of concern is a $\Gamma$ function of summation variables in the numerator, like in the factor

$$\frac{(-)^{l^j_{i',i}} M^j_{i',i}}{l^j_{i',i}!} = \frac{(-)^{k^j_{i',i}+l^j_{i',i}} \Gamma(b^j_{i',i} + l^j_{i',i} + r^j_{i',i} - k^j_{i',i})}{2 k^j_{i',i}! l^j_{i',i}!} \,. \qquad (7.1)$$

Its origin is (A.11) where the function $k_b(m|x_i - x_{i'}|)$ with a composed argument is split. Since this Bessel function is analytical, it is not dangerous in itself. Only integration in a region where the argument is zero may cause divergences.

These correspond to the ultra violet divergences in momentum space.

In the case of functions $k_b(mx)$ with simple argument, divergences were appearing through factors of the kind

$$\frac{1}{N} = \frac{1}{2v - 2\sum b + 2\sum k + \ldots}. \qquad (7.2)$$

In the case of functions with a composed argument, the divergences come from integration in the region where $|\mathbf{x}_i - \mathbf{x}_{i'}| = 0$. These manifest by divergence of the summation over the variable $l_{i',i}$. For this reason, the method of section 5 is applicable only on diagrams like fig. 4, where composed k-functions correspond to single connected vertices, which add only one term $-2b$ to the denominator of (7.2).

There is a freedom of choice of the configuration space parameters. Transformations like

$$\mathbf{x}_i \to \mathbf{x}_i + \mathbf{x}_{i'} \qquad (7.3)$$

are possible. By such transformations, it may be tried to manage that all multiple connected vertices correspond to simple k functions. This has been done in Fig. 4. However, there are cases where such transformations cannot be found, like the diagram of Fig. 7.

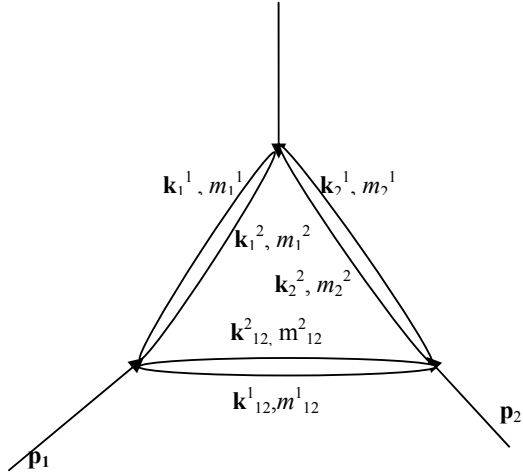

FIG. 7. A diagram with three vertices and six internal lines.

A division of the integral range, finer than in (5.4), is needed, $(V+1)!$ subregions of the kind

$$x_1 < x_2 < |\mathbf{X}_1 - \mathbf{X}_2| < X \text{ and permutations} \tag{7.4}$$

have to be considered and more complicate angle integrations have to be performed. This procedure has been discussed in Ref [14]. Fortunately, diagrams like Fig. 7 are not used in to day's physics.

A second source of divergences are integrations in the neighbourhood of $\infty$. They correspond to infra red divergences in momentum space. By caring that the asymptotic behaviour of the integrand is exponential decreasing, this kind of infinities is avoided.

In the previous sections, the above mentioned restrictions have been treated by separating out the corresponding divergences. Hence, the general expression (5.1) is convergent up to the divergences that have been separated out, and so is its expansion (5.11).

The working of the splitting procedure of (2.10) may be elucidated in the case of $\int_0^\infty dx\, x^{2\nu-1} k_\beta(x)$ with $\nu = 2$ and $\beta = 1$. The integral is known exactly:[15]

$$\int_0^\infty dx\, x^{2\nu-1} k_\beta(x) = 2^{2\nu-2} \Gamma(\nu)\Gamma(\nu-\beta) \, . \qquad (7.5)$$

On the other hand, we may work out

$$\int_0^\infty dx\, x^3 k_1(x) = \int_0^\Lambda dx\, x^3 k_1(x) + \int_\Lambda^\infty dx\, x^3 k_1(x) \, . \qquad (7.6)$$

Insertion of the expansion (A.3) into the first term of the r.h.s. and application of (D.1) yields

$$\int_0^\Lambda dx\, x^{2\nu-1} k_1(x) = \sum_{k=0}^{\infty} \frac{2^{2\nu}\{(\gamma - \frac{1}{1} - \frac{1}{2} - \frac{1}{3} \ldots - \frac{1}{k}) + (\gamma - \frac{1}{1} - \frac{1}{2} - \frac{1}{3} \ldots - \frac{1}{k} - \frac{1}{k+1})\}\left(\frac{\Lambda}{2}\right)^{2\nu+2k}}{k!(k+1)!(2\nu+2k)}$$

(7.7)

$$+ \sum_{k=0}^{\infty} \frac{2^{2\nu}\left(\frac{\Lambda}{2}\right)^{2\nu+2k} \ln\left(\frac{\Lambda}{2}\right)}{k!(k+1)!(2\nu+2k)} - \sum_{k=0}^{\infty} \frac{2^{2\nu}\left(\frac{\Lambda}{2}\right)^{2\nu+2k} \ln\left(\frac{\Lambda}{2}\right)}{k!(k+1)!(2\nu+2k)^2} + \frac{1}{2(2\nu-2)}\left(\frac{\Lambda}{2}\right)^{2\nu-2}.$$

Insertion of the asymptotic expansion (A.6a) into the second term of the r.h.s. of (7.6) yields

$$\int_\Lambda^\infty dx\, x^{2\nu-1} k_1(x) \cong \left\{\sqrt{\pi} \sum_{k=0}^{H-1} 2^{\frac{1}{2}-k}(1,k)\right\} \int_\Lambda^\infty dx\, x^{2\nu-k-\frac{5}{2}} e^{-x}, \quad (7.8)$$

on which (C.7) may be applied.

We take the number of terms of the asymptotic series 15 ($H=15$), $\gamma=0{,}5772156649$, and find

| $\Lambda$ | $\int_0^\Lambda dx\, x^3 k_1(x)$ | $\int_\Lambda^\infty dx\, x^3 k_1(x)$ | $\int_0^\Lambda dx\, x^3 k_1(x) + \int_\Lambda^\infty dx\, x^3 k_1(x)$ |
|---|---|---|---|
| 2 | 1,969921963 | 1,204499139 | 3,174421102 |
| 5 | 3,734552813 | 0,26544653 | 3,999999343 |
| 10 | 3,995697337 | 0,004301964 | 3,99999301 |
| 15 | 3,999745608 | 0,0000503234 | 3,999795931 |
| 20 | 3,951797003 | -0,0000460393 | 3,951750964 |
| 23 | 2,799347994 | -0,008824087 | 2,781523907 |

It is seen that for 10<Λ<15 the first integral alone approaches the exact answer 4 of (7.5). For smaller Λ's, the asymptotic integral may help convergence. Apparently, the calculations are not sufficiently accurate $\Lambda < 3$ for and $\Lambda > 20$.

**ACKNOWLEDGMENT**

The author thanks professor J. Smit for a useful discussion.

**APPENDIX A: BESSEL FUNCTIONS**

Most of the formulas of this appendix are found in the textbook by Watson[15] for the functions $J_\beta(z)$, $I_\beta(z)$, $K_\beta(z)$, $H^{(1)}{}_\beta(z)$ and $H^{(2)}{}_\beta(z)$. In this paper, use has been made of functions $j_\beta(z)$, $i_\beta(z)$, $k_\beta(z)$, $h^+_\beta(z)$ and $h^-_\beta(z)$ that differ from them by a factor $\left(\dfrac{z}{2}\right)^\beta$ according to

$$j_\beta(z) = \left(\frac{z}{2}\right)^{-\beta} J_\beta(z) = \sum_{k=0}^{\infty} \frac{(-)^k}{k!\,\Gamma(1+\beta+k)} \left(\frac{z}{2}\right)^{2k} \qquad (A.1a)$$

$$i_\beta(z) = \left(\frac{z}{2}\right)^{-\beta} I_\beta(z) = \sum_{k=0}^{\infty} \frac{1}{k!\Gamma(1+\beta+k)} \left(\frac{z}{2}\right)^{2k} \tag{A.1b}$$

$$k_\beta(z) = \left(\frac{z}{2}\right)^{-\beta} K_\beta(z) = \frac{\pi}{2\sin\pi\beta} \sum_{k=0}^{\infty} \left\{ \frac{1}{k!\Gamma(1-\beta+k)} \left(\frac{z}{2}\right)^{-2\beta+2k} - \frac{1}{k!\Gamma(1+\beta+k)} \left(\frac{z}{2}\right)^{2k} \right\} \tag{A.1c}$$

$$h_\beta^+(z) = \left(\frac{z}{2}\right)^{-\beta} H^{(1)}{}_\beta(z) = \frac{-i}{\sin\pi\beta} \sum_{k=0}^{\infty} (-)^k \left\{ \frac{1}{k!\Gamma(1-\beta+k)} \left(\frac{z}{2}\right)^{-2\beta+2k} - \frac{e^{-i\pi\beta}}{k!\Gamma(1+\beta+k)} \left(\frac{z}{2}\right)^{2k} \right\}$$

$$\tag{A.1d}$$

$$h_\beta^-(z) = \left(\frac{z}{2}\right)^{-\beta} H^{(2)}{}_\beta(z) = \frac{i}{\sin\pi\beta} \sum_{k=0}^{\infty} (-)^k \left\{ \frac{1}{k!\Gamma(1-\beta+k)} \left(\frac{z}{2}\right)^{-2\beta+2k} - \frac{e^{i\pi\beta}}{k!\Gamma(1+\beta+k)} \left(\frac{z}{2}\right)^{2k} \right\}.$$

$$\tag{A.1e}$$

If it is supposed that

$$\beta = b + \varepsilon, \tag{A.2}$$

where $b$ is integer, (A.1) becomes in the limit $\varepsilon \to 0$:

$$j_b(z) = \sum_{k=0}^{\infty} \frac{(-)^k}{k!(b+k)!} \left(\frac{z}{2}\right)^{2k} \tag{A.3a}$$

$$i_b(z) = j_b(iz) = \sum_{k=0}^{\infty} \frac{1}{k!(b+k)!} \left(\frac{z}{2}\right)^{2k} \tag{A.3b}$$

$$k_b(z) = K_b(k)\left(\frac{z}{2}\right)^{2k} + L_b\left(\frac{z}{2}\right)^{2k} \ln\left(\frac{z}{2}\right) + M_b(k)\left(\frac{z}{2}\right)^{-2b+2k} \quad (A.3c)$$

$$h_b^+(z) = j_b(z) + iy_b(z) \quad (A.3d)$$

$$h_b^-(z) = j_b(z) - iy_b(z), \quad (A.3e)$$

with the notation

$$y_b(z) = \frac{-(-)^k 2}{\pi}\left\{(-)^b K_b(k)\left(\frac{z}{2}\right)^{2k} + (-)^b L_b\left(\frac{z}{2}\right)^{2k} \ln\left(\frac{z}{2}\right) + M_b(k)\left(\frac{z}{2}\right)^{-2b+2k}\right\} \quad (A.4a)$$

$$K_b(k) = \frac{(-)^{b+1}}{k!(b+k)!}\left\{\left(\gamma - \tfrac{1}{1} - \tfrac{1}{2} - \ldots - \tfrac{1}{k}\right) + \left(\gamma - \tfrac{1}{1} - \tfrac{1}{2} - \ldots - \tfrac{1}{k+b}\right)\right\} \quad (A.4b)$$

($\gamma = 0{,}5772156649$),

$$L_b(k) = \frac{(-)^{b+1}}{k!(b+k)!} \quad (A.4c)$$

$$M_b(k) = \frac{(-)^k (b-k-1)!}{2 \cdot k!} \quad (A.4d)$$

From here, we see that

$$h_b^+(iz) = \frac{2i(-)^{b+1}}{\pi} k_b(z) \tag{A.5a}$$

$$h_b^-(iz) = 2i_b(z) - \frac{2i(-)^{b+1}}{\pi} k_b(z) . \tag{A.5b}$$

The asymptotic behaviour of the functions $k_\beta(z)$, $h_\beta^+(z)$, $h_\beta^-(z)$, $h_\beta^+(iz)$ and $h_\beta^-(iz)$ is given by

$$k_\beta(z) = e^{-z} \left\{ \sum_{k=0}^{H-1} \frac{\sqrt{\pi} 2^{2\beta} (\beta, k)}{(2z)^{\beta+k+\frac{1}{2}}} + O(z^{-H}) \right\} \quad (-\tfrac{3}{2}\pi < \arg z < \tfrac{3}{2}\pi) \tag{A.6a}$$

$$h_\beta^+(z) = \frac{2^{2\beta+1} e^{-\frac{1}{2}i\pi(\beta+\frac{1}{2})} e^{iz}}{\sqrt{\pi}} \left\{ \sum_{k=0}^{H-1} \frac{i^k (\beta, k)}{(2z)^{\beta+k+\frac{1}{2}}} + O(z^{-H}) \right\} \quad (-\pi < \arg z < 2\pi) \tag{A.6b}$$

$$h_\beta^-(z) = \frac{2^{2\beta+1} e^{\frac{1}{2}i\pi(\beta+\frac{1}{2})} e^{-iz}}{\sqrt{\pi}} \left\{ \sum_{k=0}^{H-1} \frac{(-i)^k (\beta, k)}{(2z)^{\beta+k+\frac{1}{2}}} + O(z^{-H}) \right\} \quad (-2\pi < \arg z < \pi) \tag{A.6c}$$

$$h_\beta^+(iz) = \frac{2^{2\beta+1} e^{-i\pi(\beta+\frac{1}{2})} e^{-z}}{\sqrt{\pi}} \left\{ \sum_{k=0}^{H-1} \frac{(\beta, k)}{(2z)^{\beta+k+\frac{1}{2}}} + O(z^{-H}) \right\} \quad (-\tfrac{3}{2}\pi < \arg z < \tfrac{3}{2}\pi) \tag{A.6d}$$

$$h_\beta^-(iz) = \frac{2^{2\beta+1} e^z}{\sqrt{\pi}} \left\{ \sum_{k=0}^{H-1} \frac{(-)^k (\beta, k)}{(2z)^{\beta+k+\frac{1}{2}}} + O(z^{-H}) \right\} \quad (-\tfrac{5}{2}\pi < \arg z < \tfrac{1}{2}\pi) \tag{A.6e}$$

$$i_\beta(z) = \frac{2^{2\beta}}{\sqrt{\pi}} \left\{ \sum_{k=0}^{H-1} \frac{(-)^k (\beta, k) e^z}{(2z)^{\beta+k+\frac{1}{2}}} + \sum_{k=0}^{H-1} \frac{(\beta, k) e^{-i\pi(\beta+\frac{1}{2})} e^{-z}}{(2z)^{\beta+k+\frac{1}{2}}} + O(z^{-H}) \right\} \quad (-\tfrac{3}{2}\pi < \arg z < \tfrac{1}{2}\pi) ,$$

$$\tag{A.6f}$$

where $(\beta,0)=1$ and

$$(\beta,k) = \frac{\Gamma(\beta+k+\frac{1}{2})}{k!\Gamma(\beta-k+\frac{1}{2})} = \frac{\{\beta^2-(\frac{1}{2})^2\}\{\beta^2-(\frac{3}{2})^2\}...\{\beta^2-(\frac{2k-1}{2})^2\}}{k!} \qquad (A.7)$$

if $k>0$.

The function $k_b(m|\mathbf{x}-\mathbf{y}|)$ may be split according to

$$k_\beta(m|\mathbf{x}-\mathbf{y}|) = \Gamma(\beta)\sum_{l=0}^{\infty}(\beta+l)C_l^\beta(\hat{\mathbf{x}}.\hat{\mathbf{y}})\left(\frac{m^2xy}{4}\right)^l i_{\beta+l}(mx)k_{\beta+l}(my) \qquad (A.8)$$

if

$$x<y, \qquad (A.9)$$

where $\hat{\mathbf{x}}.\hat{\mathbf{y}}$ is the cosine of the angle between $\mathbf{x}$ and $\mathbf{y}$ and $C_k^\beta(\hat{\mathbf{x}}.\hat{\mathbf{y}})$ are Gegenbauer polynomials, defined by

$$C_l^\beta(\hat{\mathbf{x}}.\hat{\mathbf{y}}) \equiv \sum_r{}' \frac{(-)^{\frac{l-r}{2}}\Gamma\left(\beta+\frac{l+r}{2}\right)(2\hat{\mathbf{x}}.\hat{\mathbf{y}})^r}{\Gamma(\beta)\frac{l-r}{2}!r!} . \qquad (A.10)$$

$\sum{}'$ means summation under the condition that $\frac{l-r}{2}$ is integer.

A direct splitting into Bessel functions and cosines is[8]

$$k_\beta(m|\mathbf{x}-\mathbf{y}|) = \sum_{l,r=0}^{\infty} \frac{(-)^l}{l!} \left(\frac{mx}{2}\right)^{r+2l} \left(\frac{my}{2}\right)^r \frac{(2\hat{\mathbf{x}}\cdot\hat{\mathbf{y}})^r}{r!} k_{\beta+l+r}(my) \tag{A.11}$$

if

$$x < y. \tag{A.12}$$

An integral representation is

$$k_\beta(z) = \tfrac{1}{2}\int_0^\infty d\rho\, \rho^{-\beta} e^{-\rho} e^{-\frac{z^2}{4\rho}}, \tag{A.13}$$

which establishes the connection between configuration space and Feynman parameters.[14]

An other integral representation is

$$k_\beta(z) = \frac{\sqrt{\pi}}{\Gamma(\beta+\tfrac{1}{2})} \int_1^\infty dt\, e^{-zt}(t^2-1)^{\beta-\tfrac{1}{2}}. \tag{A.14}$$

If the $k_{\tfrac{1}{2}n-1}$ transform of a function $R(p)$ is defined by

$$S(x) = 2\pi^{\tfrac{1}{2}n-1} \int_0^\infty dp\, p^{n-1} k_{\tfrac{1}{2}n-1}(px) R(p), \tag{A.15a}$$

then $R(p)$ is found by the inverse transformation[20]

$$R(p) = \frac{1}{i\pi^{\frac{1}{2}n} 2^{n-2}} \int_{c-i\infty}^{c+i\infty} dx\, x^{n-1} i_{\frac{1}{2}n-1}(px) S(x) \ . \tag{A.15b}$$

## APPENDIX B: ANGULAR INTEGRATIONS

Integration over $n$ dimensional Euclidean space of a product of cosines is done by using the formula[21,22]

$$\int d\hat{\mathbf{x}} (\hat{\mathbf{x}}.\mathbf{q}_1)^{q_1} (\hat{\mathbf{x}}.\mathbf{q}_2)^{q_2} \ldots (\hat{\mathbf{x}}.\mathbf{q}_{V-1})^{q_{V-1}} = \sum_{q_i} \frac{q_1! q_2! ..q_{V-1}!}{(q_1+q_2+..q_{V-1})!} \int d\hat{\mathbf{x}}\, \hat{\mathbf{x}}.(\mathbf{q}_1 + \mathbf{q}_2 + \ldots + \mathbf{q}_{V-1})^{q_1+q_2+\ldots q_{V-1}} \ . \tag{B.1}$$

Integration of the r.h.s. yields

$$\prod_{1<i\leq V-1} \int d\hat{\mathbf{x}}_1 \frac{(2\hat{\mathbf{p}}_1.\hat{\mathbf{x}}_1)^{q_1}}{q_1!} \frac{(2\hat{\mathbf{x}}_1.\hat{\mathbf{x}}_i)^{r_{1,i}}}{r_{1,i}!}$$

$$= \sum{}' \frac{2\pi^{\frac{1}{2}n}}{\Gamma\left(\frac{n+q_1+r_{1,2}+\ldots+r_{1,V-1}}{2}\right) \frac{q_1 - \sum_{1<a\leq V-1} q_{1,a}}{2}!} \prod_{1<i\leq V-1} \frac{(2\hat{\mathbf{p}}_1.\hat{\mathbf{x}}_i)^{q_{1,i}}}{\frac{r_{1,i} - q_{1,i} - \sum_{1<a<i} q^1_{a,i} - \sum_{i<a\leq V-1} q^1_{i,a}}{2}! q_{1,i}!}$$

$$\prod_{1\leq i'<i\leq V-1} \frac{(2\hat{\mathbf{x}}_{i'}.\hat{\mathbf{x}}_i)^{q^1_{i',i}}}{q^1_{i',i}!} \ , \tag{B.2}$$

where $\sum{}'$ means summation over the variables $q_{1,i}$ and $q^1_{i',i''}$ ($1 < i' < i'' \leq V-1$) from 0 to $\infty$ in such a way that the arguments of the faculties in the denominator are integers.

Multiple application of (B.2) yields

$$\prod_i \int d\hat{\mathbf{x}}_i \frac{(2\hat{\mathbf{p}}_i.\hat{\mathbf{x}}_i)^{q_i}}{q_i!} \frac{(2\hat{\mathbf{x}}_{i'}.\hat{\mathbf{x}}_i)^{r^j_{i',i}}}{r^j_{i',i}!} = \frac{(2\pi^{\frac{1}{2}n})^{V-1} r_{i',i}!}{\prod_j r^j_{i',i}!} A(q_i, r_{i',i}) , \qquad (B.3)$$

with notation of $\prod$ according to (5.2) and

$$r_{i',i} = \sum_c r^c_{i',i} . \qquad (B.4a)$$

The angular factor in (B.3) is given by

$$A(q_i, r_{i',i}) = \sum{}' \prod \frac{(r_{i',i} + \sum q^c_{i',i})!(q_{i',i} + \sum q_{i'}{}^c{}_i)!(2\hat{\mathbf{p}}_{i'}.\hat{\mathbf{p}}_i)^{p_{i',i}+\sum q_{i',i}{}^c}}{\Gamma\left(\frac{n+q_i+\sum r_{i,a}+\sum q^c_{i,b}+\sum q_{a,i}+\sum q_a{}^c{}_i}{2}\right)\bar{q}_i!\bar{r}_{i',i}!\bar{q}_{i',i}!r_{i',i}!q_{i',i}!p_{i',i}!q^{i'}{}_{i,i''}!q_{i'}{}^i{}_{i''}!q_{i',i}{}^{i''}!},$$
(B.4b)

where we used summation variables $q^{i'}{}_{i,i''}$, $q_{i'}{}^i{}_{i''}$ and $q_{i',i}{}^{i''}$

($1 \leq i' < i < i'' \leq V-1$) with three indices and summation notation

$$\sum r_{i,a} = \sum_{a=i+1}^{V-1} r_{i,a} , \quad \sum q_{a,i} = \sum_{a=1}^{i-1} q_{a,i} , \quad \sum q^c{}_{i,b} = \sum_{c=1}^{i-1}\sum_{b=i+1}^{V-1} q^c{}_{i,b}$$

$$\sum q_a{}^c{}_i = \sum_{a=1}^{c-1}\sum_{c=2}^{i-1} q_a{}^c{}_i , \quad \sum q_{i',i}{}^c = \sum_{c=i+1}^{V-1} q_{i',i}{}^c , \text{ etc.}$$
(B.5)

Furthermore,

$$\bar{q}_i = \frac{q_i - \sum q_{i,a} - \sum p_{a,i}}{2} \tag{B.6a}$$

$$\bar{r}_{i',i} = \frac{r_{i',i} + \sum q^a{}_{i',i} - q_{i',i} - \sum q^{i'}{}_{a,i} - \sum q^{i'}{}_{i,a}}{2} \tag{B.6b}$$

$$\bar{q}_{i',i} = \frac{q_{i',i} + \sum q_{i'}{}^a{}_i - p_{i',i} - \sum q_{a,i'}{}^i - \sum q_{i',a}{}^i}{2}. \tag{B.6c}$$

$\sum{}'$ means summation over the variables $p_{i',i}$, $q_{i',i}$, $q^{i'}{}_{i,i''}$, $q^{i}_{i',i''}$ and $q^{i}_{i',i''}$ from 0 to $\infty$ in such a way that $\bar{q}_i$, $\bar{q}_{i',i}$ and $\bar{r}_{i',i}$ are integers.

Summation over $\bar{q}_i$ yields Bessel functions:

$$\prod_i \int d\hat{\mathbf{x}}_i e^{i\mathbf{p}_i \cdot \mathbf{x}_i} \frac{(2\hat{\mathbf{x}}_{i'} \cdot \hat{\mathbf{x}}_i)^{r^j_{i',i}}}{r^j_{i',i}!} = \frac{(2\pi^{\frac{1}{2}n})^{V-1} r_{i',i}!}{\prod_j r^j_{i',i}!} \sum{}' A(r_{i',i}) \left(\frac{ip_i x_i}{2}\right)^{s_i - \sum \bar{r}_{i,a} - \sum \bar{q}_{a,i} - \sum q^i{}_{a,b} - \sum q_{a,b}{}^i} j_{\frac{n}{2}-1+s_i}(p_i x_i) \tag{B.7}$$

with

$$s_i = \frac{\sum q_{i,a} + \sum p_{a,i} + \sum r_{i,a} + \sum q^c{}_{i,a} + \sum q_{a,i} + \sum q_a{}^c{}_i}{2} \tag{B.8a}$$

$$= \bar{r}_{i,a} + 2\bar{q}_{i,a} + \bar{q}_{a,i} + p_{i,a} + p_{a,i} + q^i{}_{a,b} - q_i{}^a{}_b + q_{a,i}{}^c + q_{i,a}{}^c + q_{a,b}{}^i$$

and

$$A(r_{i',i}^j) = \prod \frac{(r_{i',i} + \sum q^c{}_{i',i})!(q_{i',i} + \sum q_{i'}{}^c{}_i)!(2\hat{\mathbf{p}}_{i'}\cdot\hat{\mathbf{p}}_i)^{p_{i',i}+\sum q_{i',i}{}^c}}{\bar{r}_{i',i}!\bar{q}_{i',i}!r_{i',i}!q_{i',i}!p_{i',i}!q^{i'}{}_{i,i''}!q_{i'}{}^i{}_{i''}!q_{i',i}{}^{i''}!} \quad (B.8b)$$

## APPENDIX C: RADIAL X→∞ INTEGRATIONS

The radial factors $R_1(r_{i',i}^j)$ and $R_2(r_{i',i}^j)$, appearing in (5.5) as result of integrations in the subregions (5.4b), may be expressed, using (A.8) and (B.8), in terms of Bessel functions. They become:

$$R_1(r_{i',i}^j) = \sum{}' \prod \frac{2^{2\nu}(-)^{\frac{l_{i',i}^j - r_{i',i}^j + l_{i,i''}^j - r_{i,i''}^j}{2}} (m_{i',i}^j)^{2l_{i',i}^j} (b_{i',i}^j + l_{i',i}^j)(b_{i,i''}^j + l_{i,i''}^j)\Gamma\left(b_{i',i}^j + \frac{l_{i',i}^j + r_{i',i}^j}{2}\right)\Gamma\left(b_{i,i''}^j + \frac{l_{i,i''}^j + r_{i,i''}^j}{2}\right)}{\left(\frac{l_{i',i}^j - r_{i',i}^j}{2}\right)!\left(\frac{l_{i,i''}^j - r_{i,i''}^j}{2}\right)!}$$

$$\times \int\limits_{0<x_1<\ldots<x_f<X} d\frac{x_i}{2}\left(\frac{ip_i x_i}{2}\right)^{s_i - \sum \bar{r}_{i,a} - \sum \bar{q}_{a,i} - \sum q^i{}_{a,b} - \sum q_{a,b}{}^i}\left(\frac{x_i}{2}\right)^{2\nu - 1 + \sum l^c_{i,a} + \sum l^c_{a,i}}$$

$$\times i_{\frac{n}{2}-1+s_i}(p_i x_i) k_{b_i^j}(m_i^j x_i) k_{b_{i',i}^j + l_{i',i}^j}(m_{i',i}^j x_i) i_{b_{i,i''}^j + l_{i,i''}^j}(m_{i,i''}^j x_i) + \text{permutations},$$

(C.1a)

which may be computed after insertion of the near zero expansions (A.3b) and (A.3c), and

$$R_2(r_{i',i}^j) = \sum {}' \prod \frac{2^{2\nu}(-)^{\frac{l_{i',i}^j - r_{i',i}^j + l_{i,i''}^j - r_{i,i''}^j}{2}} (m_{i',i}^j)^{2l_{i',i}^j} (b_{i',i}^j + l_{i',i}^j)(b_{i,i''}^j + l_{i,i''}^j) \Gamma\left(b_{i',i}^j + \frac{l_{i',i}^j + r_{i',i}^j}{2}\right) \Gamma\left(b_{i,i''}^j + \frac{l_{i,i''}^j + r_{i,i''}^j}{2}\right)}{\left(\frac{l_{i',i}^j - r_{i',i}^j}{2}\right)! \left(\frac{l_{i,i''}^j - r_{i,i''}^j}{2}\right)!}$$

$$\times \int_{X < x_{f+1} < \ldots < x_{V-1} < \infty} d\frac{x_i}{2} \left(\frac{ip_i x_i}{2}\right)^{s_i - \sum \bar{r}_{i,a} - \sum \bar{q}_{a,i} - \sum q_{a,b}^i - \sum q_{a,b}^i} \left(\frac{x_i}{2}\right)^{2\nu - 1 + \sum l_{i,a}^c + \sum l_{a,i}^c}$$

$$\times i_{\frac{n}{2} - 1 + s_i}(p_i x_i) k_{b_i^j}(m_i^j x_i) k_{b_{i',i}^j + l_{i',i}^j}(m_{i',i}^j x_i) i_{b_{i,i''}^j + l_{i,i''}^j}(m_{i,i''}^j x_i) + \text{permutations}.$$

(C.1b)

Insertion of the asymptotic expansions (A.6) into (C.1b) yields incomplete $\Gamma$ functions. The power of $x_i$ is restricted since terms $s_i$ and $l_{i',i}^j$ in the exponent are cancelled by terms $s_i$ and $l_{i',i}^j$ in the order of the Bessel functions.

(C.1b) is computed by use of the following formula's:

$$\int_X^\infty dx_1 e^{-M_1 x} x_1^{\rho_1 - 1}$$

$$= M^{-\rho_1} \sum_{0 \le k_1 \le H_1 - 1} \frac{\Gamma(\rho_1)}{\Gamma(\rho_1 - k_1)} e^{-M_1 X} (M_1 X)^{-k_1 - 1} + M^{-H_1 - 1} \frac{\Gamma(\rho_1)}{\Gamma(\rho_1 - H_1)} \int_X^\infty dx_1 e^{-M_1 x} x_1^{\rho_1 - H_1 - 1}.$$

(C.2)

If $\rho_1$ is positive integer, the series in the r.h.s. is finite. If $\rho_1$ is positive non integer, $H_1$ is chosen $\ge \rho_1$ so that the exponent of $x_1$ in the integrand is negative.

In both cases and also in the case $\rho_1 \leq 0$ the r.h.s is restricted according to

$$\int_X^\infty dx_1 e^{-M_1 x} x_1^{\rho_1-1} \leq M_1^{-\rho_1} \sum_{0 \leq k_1 \leq H_1} \frac{\Gamma(\rho_1)}{\Gamma(\rho_1 - k_1)} \Lambda^{\rho_1 - k_1 - 1} e^{-\Lambda} , \qquad (C.3)$$

where

$$\Lambda = M_1 X . \qquad (C.4)$$

Multiple application of this formula yields

$$\prod_{1 \leq i \leq g} \int_{X < x_1 \ldots < x_g < \infty} dx_i\, e^{-M_i x} x_1^{\rho_i - 1} \leq \left(\sum_{1 \leq a \leq g} M_a\right)^{-\sum_{1 \leq a \leq g} \rho_a} \sum_{0 \leq k_i \leq H_i} \frac{\Gamma(\sum_{i \leq a \leq g} \rho_a - \sum_{i < a \leq g} k_a - g + i)}{\Gamma(\sum_{i \leq a \leq g}(\rho_a - k_a) - g + i)} \left(\frac{\sum_{1 \leq a \leq g} M_a}{\sum_{i \leq a \leq g} M_a}\right)^{-1-k_i} \Lambda^{\sum_{1 \leq a \leq g} \rho_a - k_a - g} e^{-\Lambda}. \qquad (C.5)$$

It is seen that (C.3) and (C.5) can be made arbitrary small by choosing $\Lambda$ sufficiently large.

By application of

$$\int_X^\infty dx_1 e^{-M_1 x_1} x_1^{\rho_1-1} = M_1^{\rho_1} \Gamma(\rho_1) - \int_0^X dx_1 e^{-M_1 x_1} x_1^{\rho_1-1} , \qquad (C.6)$$

the incomplete Γ-functions with integer or half integer arguments may be computed by the following expansions:

$$\int_X^\infty dx_1 e^{-M_1 x} x_1^{\rho_1-1} = P_{\rho_1} + Q_{\rho_1} \sum_{\substack{l=0,\\ l\neq -\rho_1}}^\infty \frac{(-)^l \Lambda^{\rho_1+l}}{l!(\rho_1+l)} ,$$ (C.7)

where

$$P_{\rho_1} = (\rho_1-1)! \text{ and } Q_{\rho_1} = -1 \text{ if } \rho_1 \text{ positive integer},$$ (C.8a)

$$P_{\rho_1} = 0 \text{ and } Q_{\rho_1} = 1 \text{ if } \rho_1 \text{ non positive integer},$$ (C.8b)

$$P_{\rho_1} = \frac{\sqrt{\pi}(2\rho_1-1)!}{(\rho_1-\frac{1}{2})! 2^{2\rho_1-1}} \text{ and } Q_{\rho_1} = -1 \text{ if } \rho_1 \text{ positive half integer}$$

(C.8c)

$$P_{\rho_1} = \frac{\sqrt{\pi}(-)^{\frac{1}{2}-r}(\frac{1}{2}-\rho_1)! 2^{1-2\rho_1}}{(1-2\rho_1)!} \text{ and } Q_{\rho_1} = -1 \text{ if } \rho_1 \text{ negative half integer}$$

(C.8d)

Multiple application of (C.7),(C.8) yields

$$\prod_{1\le i\le g} \int_{X<x_1<...<x_g}^\infty dx_i\, e^{-M_1 x} x_i^{\rho_i-1} = \prod_{1\le i\le g} \int_{X<x_1<...<x_g}^\infty dx_i\, e^{-M_1 x} x_i^{\rho_i-1} = P_{\rho_1}...P_{\rho_g} M_1^{-\rho_1}....M_g^{-\rho_g}$$

$$+ P_{\rho_1}...P_{\rho_{g-1}+\rho_g+l_g} Q_{\rho_g} M_1^{-\rho_1}...M_{g-1}^{-\rho_{g-1}-\rho_g} \frac{\left(-\frac{M_g}{M_{g-1}}\right)^{l_g}}{l_g!(\rho_g+l_g)}$$

$$+ 2^g - 1 \quad \text{terms where factors } P_{\rho_i} \text{ have been replaced by factors } Q_{\rho_i}.$$
(C.9)

Equation (C.9) may be inserted into (C.1b).

# APPENDIX D: RADIAL 0→Λ INTEGRATIONS

Multiple application of

$$\prod_j \int_0^\Lambda d\xi_1 \xi_1^{N_1-1} \left\{\ln(\mu_1^j \xi_1)\right\}^{g_1^j}$$

$$= \sum_{0 \leq g_{1;1}^j \leq g_1^j} \prod_j \frac{g_1^j!(-)^{\sum_c (g_1^c - g_{1;1}^c)} \sum_c (g_1^c - g_{1;1}^c)!}{g_{1;1}^j!(g_1^j - g_{1;1}^j)! N_1^{1+\sum_c (g_1^c - g_{1;1}^c)}} \left\{\ln(\mu_1^j \Lambda)\right\}^{g_{1;1}^j} \Lambda^{N_1}. \tag{D.1}$$

gives:

$$\prod \int_{0<\xi_1<\xi_2<..<\xi_{V-1}<\Lambda} d\xi_i\ \xi_i^{N_i-1}\left(\ln \mu_i^j \xi_i\right)^{g_i^j}$$

$$= \sum_{0 \leq f_i^j \leq g_{i;V-2}^j \ldots \leq g_{i;i}^j \leq g_i^j} \prod_{1 \leq i \leq i'' \leq V-1} \frac{(-)^{G_i} G_i! g_i^j!}{(g_{i;i''-1}^j - g_{i;i''}^j)! \left(\sum_{1 \leq a \leq i} N_a\right)^{1+G_i}} (\ln \mu_i^j \Lambda)^{f_i^j} \Lambda^{\left(\sum_{1 \leq a \leq V-1} N_a\right)} \tag{D.2}$$

with

$$G_i = \sum_{\substack{1 \leq a \leq i \\ c}} (g_{a;i-1}^c - g_{a;i}^c) . \tag{D.3}$$

$g_{i;i''}^j$ are summation variables if $i \leq i'' \leq V-1$ and

$$g_{i;i-1}^j \equiv g_i^j \tag{D.4a}$$

$$g^j_{i;V-1} \equiv f^j_i . \qquad (D.4b)$$

In the case

$$0 \leq g^j_i \leq 1 , \qquad (D.5)$$

all summation variables are 0 or 1 and (D.2) takes the form

$$\prod_{0<\xi_1<\xi_2<..<\xi_{V-1}<\Lambda} \int d\xi_i \ \xi_i^{N_i-1} (\ln \mu^j_i \xi_i)^{g^j_i}$$

$$= \sum_{0 \leq f^j_i \leq g^j_{i;V-2}...\leq g^j_{i,i} \leq g^j_i} \prod_{\substack{1 \leq i \leq V-1 \\ j}} \frac{(-)^{G_i} G_i!}{\left(\sum_{1 \leq a \leq i} N_a\right)^{1+G_i}} \left(\ln \mu^j_i \Lambda\right)^{f^j_i} \Lambda^{\left(\sum_{1 \leq a \leq V-1} N_a\right)} . \qquad (D.6)$$